Rene Peter Bremm*, Andreas Werle, Christoph Auer, Frank Hertel, Jorge Gonçalves, Klaus Peter Koch

# TreCap: A wearable device to measure and assess tremor data of visually guided hand movements in real time

**Abstract:** The assessment and treatment of motor symptoms such as tremor in Parkinson's disease depends exclusively on the physician's visual observation of standardised movements (i.e. motor tasks). Wearable sensors such as accelerometers are able to detect some manifestations of these pathological signs in movement disorders. Sensor data from motor tasks, however, must be processed sequentially with annotated data from clinical experts. Hence, we designed TreCap, a custom-built wearable device with new software to capture and evaluate motor symptoms such as tremor in real time. Inertial sensor data is systematically processed, stored and tailored to each motor task by this software, including annotated data from clinical rating scores and deep brain stimulation parameters. For prototype testing, the wearable device was validated in a pilot study on subjects with physiological hand tremor. The processed data sets are suitable for machine learning to classify motor tasks. Results on healthy subjects demonstrate an accuracy of 95% with support vector machine algorithms. TreCap software is expandable and allows full access to the configuration of all sensors via Bluetooth®. Finally, the functions of the entire device provide a platform to be apt for future clinical trials.

# Introduction

Wearable sensors offer significant opportunities to improve personalised medicine and to develop new digital health applications [1]. For example, using smart devices (phones, watches, custom-built research systems) for disease diagnostics and monitoring of movement disorders such as in Parkinson's disease (PD) and essential tremor (ET) [2], [3], [4]. Both diseases affect the lives of a large number of patients worldwide [5]. Tremor is a common motor symptom in movement disorders and occurs mainly on the upper limbs [6]. Diagnosis and treatment, however, often poses a challenge for the physician [7]. In early stages of the disease, pharmaceutical treatments can be effective in suppressing tremor. Unfortunately, these treatments often tend to lose their effectiveness over time, with severe motor and psychological consequences to everyday life [8]. Deep brain stimulation (DBS) is an established method for the treatment of movement disorders [9]. It has become a standard procedure, now safely and routinely conducted worldwide, including patients with early motor complications in PD [10].

### A. Clinical Assessment

The severity of motor symptoms in movement disorders is assessed by the examination of a series of visually guided hand movements (i.e. motor tasks) according to established rating scales such as the *Unified Parkinson's Disease Rating Scale*

---

*****Corresponding author: Rene Peter Bremm**, Luxembourg Centre for Systems Biomedicine, Interventional Neuroscience Group, University of Luxembourg, Esch-sur-Alzette, Luxembourg, and National Department of Neurosurgery, Centre Hospitalier de Luxembourg, Luxembourg, E-Mail: bremmrp@outlook.com, https://orcid.org/0000-0002-6782-4026
**Andreas Werle:** Department of Electrical Engineering, Trier University of Applied Sciences, Trier, Germany.
**Christoph Auer:** Department of Electrical Engineering, Trier University of Applied Sciences, Trier, Germany.
**Frank Hertel:** National Department of Neurosurgery, Centre Hospitalier de Luxembourg, Luxembourg; and Interventional Neuroscience Group, Luxembourg Centre for Systems Biomedicine, University of Luxembourg, Esch-sur-Alzette, Luxembourg.
**Jorge Gonçalves:** Systems Control Group, Luxembourg Centre for Systems Biomedicine, University of Luxembourg, Esch-sur-Alzette, Luxembourg.
**Klaus Peter Koch:** Department of Electrical Engineering, Trier University of Applied Sciences, Trier, Germany, E-Mail: koch@hochschule-trier.de

(UPDRS) [11] and *The Essential Tremor Rating Assessment Scale* (TETRAS) [12]. This standardised neurological examination partially defines the clinical condition of a patient and is used to describe the impact on functional ability and quality of life [13]. However, an accurate estimation of motor impairments is essentially dependent on the observational skills of the physician, who often prefer objective tools to improve their work [7].

Tremor occurs in healthy individuals and as a symptom of movement disorders. The oscillation often increases with mental stress, fatigue or emotions [14] and is mostly described by its characteristics in terms of frequency, amplitude and regularity [15]. Physiological hand tremor (9-13 Hz) is characterized by low amplitude and high frequency components. This kind of tremor has a wide range of frequencies with distinct time variability [16]. In a previous study, approximately 8 % of young and elderly adults were found to have a physiological hand tremor acceleration pattern that is an indistinguishable form of mild ET [16]. This observation provides a framework for the interpretation of physiological studies in subjects with unsuspected pathological tremor [17].

Pathological tremor represents an involuntary movement with a relatively fixed tremor frequency. Its amplitude shows short- and long-term variability, which is influenced by the progress of the disease and the effectiveness of the treatment. The pathological oscillation is classified into rest tremor or action tremor [11]. PD tremor (3-6 Hz) typically occurs at rest. ET (2-7 Hz) is mainly characterised by an action tremor, which is further divided into postural and kinetic tremors [6]. A subtype of the kinetic tremor is the intention tremor, which occurs mainly in multiple sclerosis [18] and ET patients with mixed tremor syndromes [19]. This tremor increases towards the end of a visually-guided goal-directed movement [20]. A clinical hand movement to evaluate this task-specific tremor is the finger to nose (FN) motor task [21].

### B. Device-based assessment

Previous studies have shown that inertial sensors provide reliable information on physical activities [22] and have also been accepted as useful tools in a clinical research facility [23], [24]. Their limitations, potentials and current progress is reported in several studies [25], [26], [27], [28], [29]. For instance, acceleration sensors are able to detect some manifestations of pathological signs in PD, such as tremor, bradykinesia, dyskinesia, and freezing of gait [30], [31]. In addition, accelerometers are able to detect minute variations, which are often imperceptible during visual examination by a physician [32]. This information may help to distinguish a patient from a healthy subject in the early stages of the disease.

Several authors have contributed to signal processing, analysis and classification methods of accelerometer data in a manner suitable for the extraction of clinical parameters such as for tremor in PD [32], [33], [34]. Spectral analysis of pathological movements and the analysis of waveform properties often refines the visual impression from clinical observations [35], [36]. The use of accelerometers alone, however, does not capture the energy cost of certain activities [37]. Physical activities may be underestimated depending upon orientation, position, and velocity. Hence, an inertial measurement unit (IMU) is often used by means of an accelerometer, gyroscope, and magnetometer [38]. In addition, neuromuscular activity using electromyography (EMG) provides additional information [39], [40]. Hence, a combination of kinematic and bioelectrical measurements can help to quantify the effects of DBS, anti-Parkinsonian medication, or other treatments on motor symptoms more precisely. The interpretation of sensor data, however, is often highly technical and includes extensive staff time to process and analyse the data [7]. Therefore, it is important to simplify sensor information and transfer it to routine clinical applications.

### C. TreCap device

This work presents *TreCap* (TremorCapture) a custom-built wearable device with novel software to evaluate and manage tremor data of visually guided hand movements in real time. Inertial sensors measure limb kinematics and a bioelectrical unit is able to measure neuromuscular activity via surface EMG. TreCap software offers systematic processing and pre-analysis of bioelectric and kinematic signals. Its graphical user interface (GUI) is mainly designed for clinical researchers and experts in movement disorders with a research background. TreCap gives all potential users the opportunity to process and manage tremor data in a standardised procedure. Recent studies demonstrate that objective measurements in routine care of patients with movement disorders can improve clinical outcomes [7], [41]. Compared to other devices, TreCap is optimised for movement disorder studies in a clinical setting and has several new practical functions [4], [42], [43]. For instance, sensor data from motor tasks are immediately processed by simply pressing a view button on an intuitive GUI. In addition, the software provides access to the raw data and configuration of all sensors. In this paper we describe how the TreCap system is structured, as well as its validation and performance.

## Material and Methods

TreCap is a custom-built device composed of a sensor prototype and software application. Figure 1 shows the wearable device mounted with an adjustable wristband on the forearm of the subject. The hardware device consists of inertial sensors to measure acceleration, angular velocity, and magnetic field. Additionally, a bioelectrical unit is integrated for future measurements of neuromuscular activity via EMG. Sensor data is transmitted via Bluetooth® radio device to a custom-written MATLAB® software. A microSD card on the hardware device offers the capacity to ensure data backups (log and stream data) and external measurements (log data), such as during the stationary stay of patients in the hospital. The design and fabrication processes of the TreCap sensor and software are summarised here. All crucial components, whether software, hardware, firmware, or sensor housing, were developed in-house. This work was first introduced in [44].

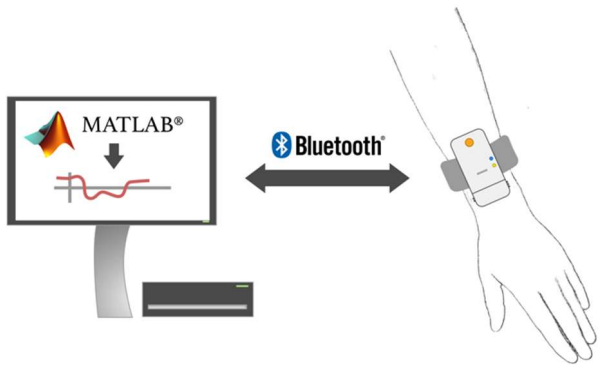

**Figure 1:** Design of the TreCap system.

### A. Sensor design and fabrication

The prototype consists of inertial sensors and a unit for sensing bioelectrical signals. Figure 2 shows the printed circuit board (PCB), which is mainly assembled with a microcontroller, an IMU, an analogue-to-digital converter (ADC), a Bluetooth® radio device, and a user interface. A lithium polymer (LiPo) battery is embedded in a housing that is fabricated by 3D rapid prototyping technology. The size of the sensor housing in Figure 1 is 75 mm × 35 mm × 25 mm. With all components such as PCB, battery, and housing, the sensor weighs 48 g. The components of the PCB were chosen to be suitable for the clinical research in movement disorders.

To measure limb kinematics, a 6+3 degree of freedom (DoF) low-power and high-performance IMU is embedded on the PCB. The 6-DoF inertial sensor is fabricated on a monolithic three-axis accelerometer plus three-axis gyroscope (Maxim Integrated™, MAX21100). This unit is connected to a three-axis magnetometer sensor (STMicroelectronis Corp., LIS3MDL). All inertial sensors provide a digital output of 16-bit resolution with an axis alignment referred to the 6+3-DoF IMU. The output data rate (ODR) and sampling frequency of the IMU can be configured according to Table 1. The TreCap software enables the configuration of the TreCap sensor via the Bluetooth® interface. For this purpose, the user uploads a configuration file in which the sensor registers are previously entered. The default settings of the device are configured for the examination of motor symptoms such as tremor on the upper limbs. The low-noise mode given in Table 1 is used and a sampling frequency of 200 Hz (samples per second, SPS) is selected for standardisation reasons [45].

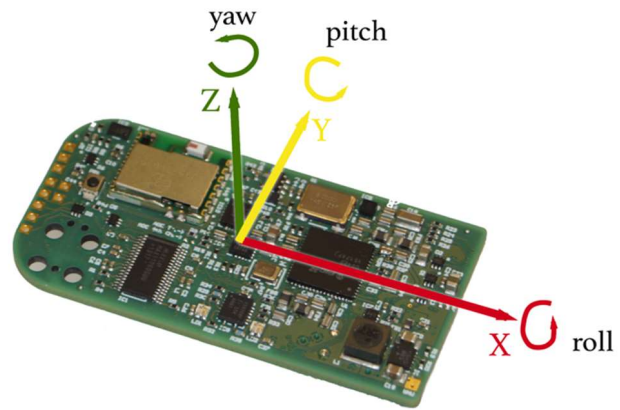

**Figure 2:** PCB with axis arrangement of the IMU.

In addition, an ADC (Maxim Integrated™, MAX11060) is integrated on the PCB to measure neuromuscular activity. This group of sigma-delta ADC provides 16-bit resolution and four modulators, which simultaneously convert each fully differential analogue input channel with a programmable ODR ranging from 250 SPS to 64 kSPS. The ADC is used with the specific goal of recording high-quality surface EMG signals. For this, an EMGs signal amplification and filtering circuit is in development using an instrumentation amplifier. This key architecture is conceived as an external module via a detachable connection to the ADC.

The hardware on the PCB is controlled by a PIC32MZ microcontroller (Microchip Inc., PIC32MZ2048EFH064) and its associated firmware is developed with MPLAB® Harmony (Microchip Inc.). The general term software in this work refers to TreCap's personal computer application. In this powerful framework, a fully integrated firmware development platform provides the drivers for the peripherals and middleware components, such as the support for the real-time operating system freeRTOS™. It has been ported to the PIC32MZ microcontroller family and is distributed under the MIT

**Table 1:** Technical specifications of the IMU.

| IMU | Accelerometer | Gyroscope | Magnetometer |
| --- | --- | --- | --- |
| **Sensor type** | MAX21100 (16-bit) | MAX21100 (16-bit) | LIS3MDL (16-bit) |
| **Sensor range** | ±2 g, ±4 g, ±8 g, ±16 g | ±250 dps, ±500 dps, ±1000 dps, ±2000 dps | ±4 G, ±8 G, ±12 G, ±16 G |
| **Output data rate (ODR, in Hz)** | Low noise mode:<br>31.25 – 2000 (min, max)<br>Eco mode:<br>0.98 – 250 (min, max) | Low noise mode:<br>3.9 – 8000 (min, max)<br>Eco mode:<br>31.25 – 250 (min, max) | Eco mode:<br>80 (typ)<br>Fast mode:<br>up to 1000 (max) |
| **Noise density (25 °C)** | Low noise mode:<br>140 µg/√Hz (typ)<br>260 µg/√Hz (max)<br>Eco mode:<br>800 µg/√Hz (ODR = 250 Hz) | Low noise mode:<br>0.009 dps/√Hz (typ)<br>0.025 dps/√Hz (max)<br>Eco mode:<br>0.018 dps/√Hz (ODR = 250 Hz) | RMS noise:<br>(±12 G range)<br>X = 3.2 mG<br>Y = 3.2 mG<br>Z = 4.1 mG |
| **Sensitivity** | 15 digit/mg (±2 g)<br>7.5 digit/mg (±4 g)<br>3.75 digit/mg (±8 g)<br>1.875 digit/mg (±16 g) | 15 digit/dps (±2000 dps)<br>30 digit/dps (±1000 dps)<br>60 digit/dps (±500 dps)<br>120 digit/dps (±250 dps) | 6842 LSB/G (±4 G)<br>3421 LSB/G (±8 G)<br>2281 LSB/G (±12 G)<br>1711 LSB/G (±16 G) |
| **Nonlinearity (25 °C, in %)** | 0.5 – 1.2 (range ±2 g, best fit line) | 0.4 (range ±2000 dps, best fit line) | ±0.12 (range ±12 G, best fit line) |
| **Cross axis (%)** | -5 / ±1 / 5 (min / typ / max) | -5 / ±1 / 5 (min / typ / max) | not available |

Sensor range is described by minimum and maximum values of acceleration, angular velocity, and magnetic field. The bandwidth of each inertial sensor is defined as the highest frequency signal that can be sampled without aliasing by the specified output data rate (ODR). Noise density is mainly defined as RMS noise or PSD per √Hz, such as dps/√Hz (degrees per second, dps). Sensitivity is described as a ratio of the sensors' electrical output to mechanical input. Sensor nonlinearity describes the deviation from a perfectly constant sensitivity, with respect to the full-scale range. Cross-axis sensitivity measures how much the output is seen on one axis when the signal is imposed on a different axis [46], [47].

License. Figure 3 depicts the communication architecture on the PCB. The basic functionality of the firmware is validated for each communication process between the microcontroller and its corresponding components via the system bus on the PCB (SPI, UART, I2C).

TreCap firmware transmits data wirelessly to the software application in real time by using a 2.4 GHz Bluetooth® radio device (STMicroelectronics Corp., SPBT2632C2A). Raw and processed data are stored in the TreCap software via an accessible swap file. The firmware is able to log data on the microSD card. Hence, a microSD card connector (Molex Corp., 503182-1853) is embedded on the PCB. Power management and user interface are handled by the microcontroller. The hardware user interface consists of one control button (Omron Corp., B3U-1100P) and two light-emitting diodes (LEDs) for TreCap's on-board control and status indication. Via the control button, the idle state (sleep mode) of the PCB is enabled or disabled.

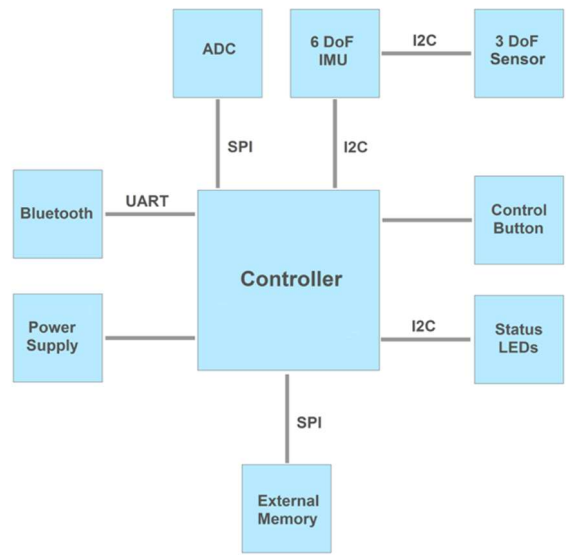

**Figure 3:** Communication architecture of the PCB.

The power supply of on-board components is ensured by a LiPo battery (Ehighsen Technology Ltd., PL703048) with a nominal voltage and capacity of 3.7 V and 1050 mAh in combination with a buck-boost converter with an output voltage of 3.3 V (STMicroelectronics Corp., STBB1-APUR). The selected rechargeable battery is also assembled with a separated protection circuit for over-voltage, over-current, and depth discharge. To ensure that the TreCap sensor operates properly, battery voltage is continuously monitored by the firmware to enable the sleep mode on the PCB when the LiPo voltage is below 3 V.

**B. Software design and implementation**

TreCap's personal computer software is based on a custom-written MATLAB® and Java™ code. Figure 4 illustrates the software architecture. For the purpose of clinical testing, a GUI is integrated and built up as a software wizard of a sequence of five dialog windows. The dialog windows one to four store metadata about the patient (pseudo code and disease), sensor setup (type and placement), measurement procedure (medication on/off and/or DBS programming), and DBS device (pulse generator, electrodes, and initial settings).

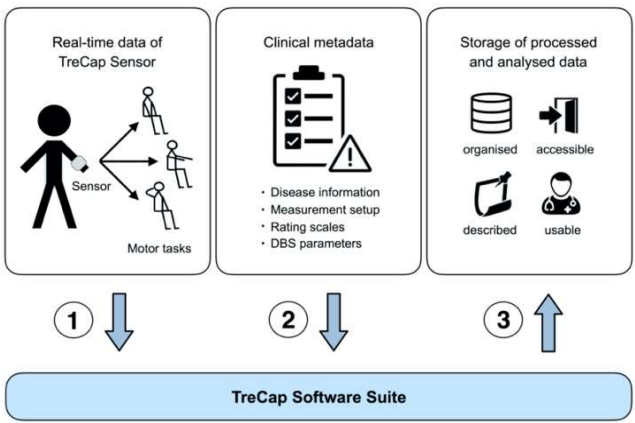

**Figure 4:** Input/output structure of the TreCap software. (1) Real-time sensor data is systematically processed and tailored to each visually guided hand movement (i.e. motor task) by a GUI, including (2) metadata such as clinical rating scores. (3) The semi-automated software tool reduces the time required for data management and post-processing.

Figure 5 shows the main (fifth) dialog window of the *TreCap Software Suite*. The GUI is divided into user-defined panels, including user checkboxes, edit text boxes and drop-down menus. Here, the upper half part of the GUI contains panels to *connect/disconnect* the sensor to the software via Bluetooth® (toggle button), *start/stop* the measurement, and *enable/disable* motor tasks with the associated metadata. The modular design of the software allows to add further motor tasks.

In addition, sensor data of the motor tasks are assigned to *rating scores* via the integrated buttons in Figure 5. The rating scale ranges from grade 0 to 4 with score 0 for absent, 1 for slight, 2 for mild, 3 for moderate, or 4 for severe tremor. Scoring 0 to 4 corresponds to TETRAS and the UPDRS motor examination part III [11], [12]. These are simple annotations performed by the physician when pressing the buttons in the GUI. There is no automatic segmentation of hand movements, no identification of motor symptoms and no prediction of the severity of symptoms. The GUI provides the platform to implement these features in the future. However, the functions in the GUI simplify real-time data processing and management as shown in Figure 4, and the visualization of the sensor signals allows the user to interpret the tremor waveform.

Moreover, a *motor task data sequence* is assigned to DBS parameter information such as amplitude, frequency and pulse width when the *DBS checkbox* is activated. The drop-down menu of each stimulation parameter displays the available steps of change based on the most common clinician programmer devices [48]. The following settings can be selected: *Amplitude* in steps of 0.1 or 0.5 mA (range 0.1 to 20 mA), *frequency* in steps of 1, 2, 5 or 10 Hz (range 2 to 255 Hz), and *pulse width* in steps of 10 μs (range 10 to 450 μs). For instance, select a motor task and increase/decrease via the *add/subtract button* the amplitude with 0.1 or 0.5 mA to the desired value. Initial DBS parameters in the grey text boxes are then updated until the final values are confirmed by pressing the *set button*. The latter ensures the assignment of the metadata (i.e. DBS parameter information) to the *motor task data sequence* and to display the information in the *event list* box. This option supports the evaluation of motor symptoms and postprocessing of motor task data in patients with DBS therapy. In addition, a marker is assigned to the *motor task data sequence* when the side effect *(SE) button* is pressed. The associated drop-down menu in the *notes panel* has the following options: None, muscle cramps, paraesthesia, headache, dyskinesia, speech impairment, visual complaint, and cognitive impairment. To identify when an optimal therapeutic effect is reached in DBS programming, a marker is assigned to the *motor task data sequence* by pressing the set-point *(SP) button*.

The lower half part of the GUI in Figure 5 shows the streaming panel, including signal pre-processing functions. Real-time signals, such as acceleration, angular velocity and magnetic field are visualised in a strip chart. To filter visually guided hand movements, a two-pole digital Chebyshev filter (high, low, and band-pass) is implemented in the custom-written MATLAB® code. The filter is available for all sensors and is applied to the selected axes when the *filter checkbox* is enabled.

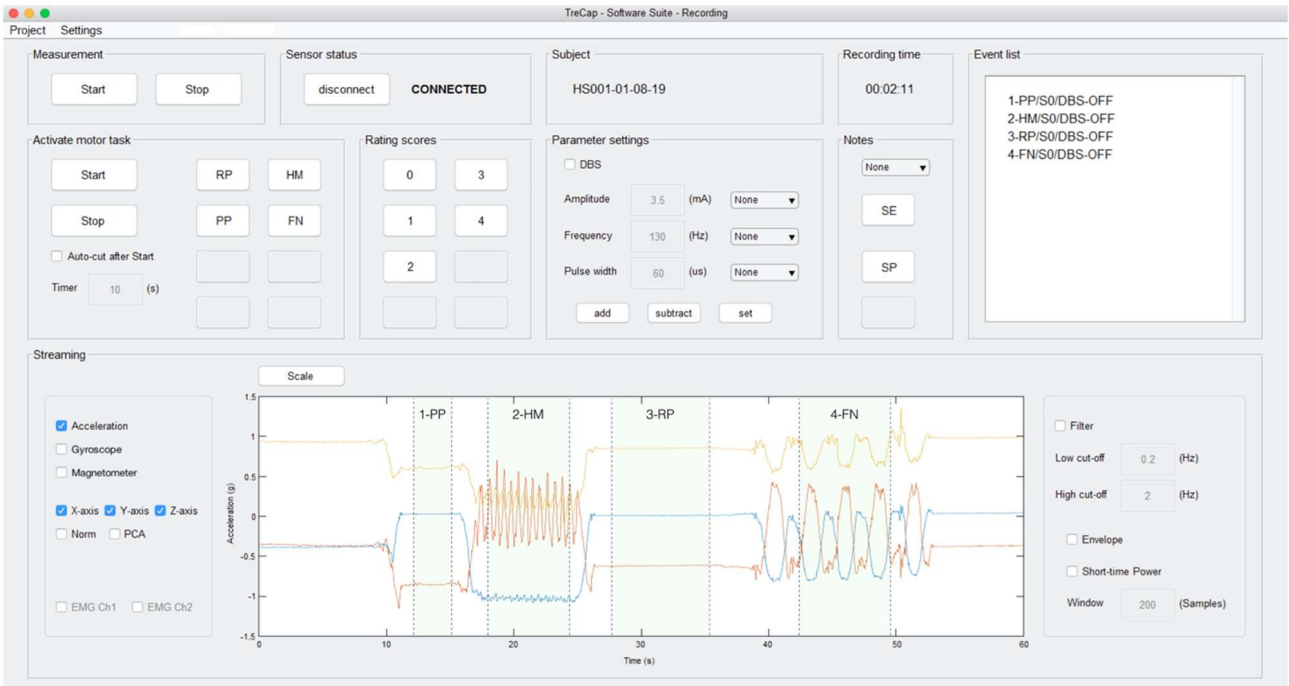

**Figure 5:** Screenshot of the GUI for the monitoring and assessment of visually guided hand movements in healthy subjects and patients with movement disorders. As an example, motor tasks of a healthy subject in rest position (RP), posture position (PP), finger to nose (FN) and hand movement (HM) are shown in the strip chart (accelerometer, X-axis in blue, Y-axis in red, Z-axis in yellow) and event list box. In this example, the Pronation/Supination movement for HM is displayed. Physiological hand tremor was measured in PP (see event list box, motor task 1). DBS parameter adaptations appear in the event list box as, for example, 5-FN/S3/DBS-3.5V-130Hz-60μs/SE6 and is to read as motor task 5, score 3, DBS settings 3.5 V, 130 Hz, 60 μs, speech impairment as side effect (SE, in notes panel, option six in drop-down menu).

Applications for signal pre-processing are applied to the accelerometer data in real time and include the visualisation of the principal component scores (via principal component analysis, PCA), the signal norm, the signal envelope (via Hilbert transform [49]), and the short-term power of the time series data. This offers several combinations together with the digital filter. All data processing routines and GUI features were implemented on the basis of our previous clinical studies [50], [51]. The methodological relevance and efficacy will be part of a future clinical study. Some examples are reported in the literature [52]. The computation of the short-term power of the time series data is a new feature.

Sensor and metadata are stored in a log file, including a pre-analysis of the processed *motor task data sequence* by the TreCap software. The pre-analysis is based on statistical metrics in the time and frequency domain, as described in the next section (prototype testing). In addition, an analysis software is under development with the specific aim of creating automated reports of the processed data.

### C. Prototype testing
C1. <u>Sensor validation</u>:
In order to design a meaningful evaluation platform to benchmark the performance of the TreCap device, the experimental verification of the inertial sensors is based on a noise and motion analysis. Due to the difficulty of functional testing, as it applies to various engineering products in industry, TreCap was compared to a commercial sensor (CS) [53] with similar IMU specifications (InvenSense Corp., MPU-9150) [54]. The CS does not provide a software application. Raw data is read out from the internal NAND flash memory via USB using Python™ code. The integrated IMU of the CS is widely used in smartphones/-watches. Previous studies report on the analysis of PD motor symptoms with this IMU [2], [3]. However, access to the raw data on

smartphones/-watches is often denied by the software and outstanding issues remain around the energy constraints when processing large data sets [55]. Hence, we first measured some basic features with each of the sensor IMUs. These features were compared to the specifications of the data sheet. Second, we validated the software under real conditions.

The IMU on the PCB measures 6+3 DoF signals with the output signal

$$s[n] = \begin{bmatrix} s_a[n] \\ s_\omega[n] \\ s_\phi[n] \end{bmatrix} = \begin{bmatrix} a[n] \\ \omega[n] \\ \phi[n] \end{bmatrix} + \begin{bmatrix} \xi_a[n] \\ \xi_\omega[n] \\ \xi_\phi[n] \end{bmatrix}, \quad (1)$$

which is expressed as the sum of the response of the inertial sensors and a noise term. In equation (1), $s[n]$ is composed of the accelerometer $s_a[n]$, gyroscope $s_\omega[n]$, and magnetometer $s_\phi[n]$ outputs, where $n$ represents the temporal index of the signal. For example, $a[n] + \xi_a[n]$ is the raw acceleration vector plus the noise vector associated with that output. For prototype testing, we measured sensor noise and sensor motion. For example, the gyroscope and magnetometer were tested in motion with a rocking shaker (Heidolph™, Platform Shakers, Duomax 1030), confirming its accuracy on the data sheet. For illustration purposes, sensor evaluation results presented in this paper focus mainly on accelerometer data in $s_a[n] \in \mathbb{R}^3$. The described statistical metrics apply to all inertial sensors in $s[n]$.

To understand how sensor noise is represented in $s_a[n]$, the TreCap sensor was mounted onto a pneumatic vibration isolation workstation (VIW; Newport™, VH3048W-OPT). The VIW reduces transmitted vertical and horizontal vibrations, making it an ideal working platform for vibration-influenced devices such as accelerometers. In equation (1), $a[n]$ is zero for noise measurements and that the environmental noise $\xi_a$ is mainly reduced by the use of the VIW. To test the accelerometer in motion, a small vibration exciter (Brüel & Kjaer™, Vibration Exciter, Type 4809) was used, which is capable of delivering peak sine forces up to 45 N. The vibration exciter features a wide frequency range (10 Hz to 20 kHz) and a continuous displacement of 8 mm peak-to-peak. This benchtop unit offers dependability for a range of applications, including accelerometer calibration. Type 4809 was driven by a custom-built amplifier and waveform generator (Keysight™, Waveform generator, 33500B Series) with a sine excitation frequency of 10 Hz. The amplitude of the generator was set to obtain an output displacement with TreCap of 5 m/s² (0.51 g) and 10 m/s² (1.02 g, sine wave, peak-to-peak). The output signal of the vibration exciter was then amplified using a small charge amplifier (Brüel & Kjaer™, Charger amplifier, Type 2635) and routed to a digital oscilloscope (Rohde & Schwarz™, Digital oscilloscope, RTB2004 Series). Type 2635 was configured with a charge amplification of factor 10 and a lower frequency limit of 2 Hz as high-pass filter.

There are two types of noise in accelerometers—electronic noise from the application-specific integrated circuit (ASIC) and mechanical noise from the *micro-electromechanical system* (MEMS) [56]. The latter is thermo-mechanical noise of the moving parts in the MEMS g-cell. To estimate the overall system noise, denoted by $w$, the mean value $\bar{s}_a[n]$ was subtracted from $s_a[n]$. The root-mean square (RMS) of the detrended noise signal $w$ was then calculated for each accelerometer axis ($a_X$, $a_Y$, $a_Z$) according to

$$N_{RMS} = \sqrt{\frac{1}{N}\sum_{n=1}^{N} w_n^2}, \quad (2)$$

where $N_{RMS}$ is the RMS for $w$ and $N$ is the number of sampled data points [57]. Here, the result for each sensor axis was compared to the CS. The $N_{RMS}$ in equation (2) is equal to the standard deviation of $w$. In addition, the normalised autocorrelation was computed to demonstrate that the noise is predominantly considered as Gaussian white noise with zero-mean. In the data sheets of the IMUs, power spectral density (PSD) is often used to describe the overall system noise and how the power is distributed over frequency. The PSD noise was calculated as

$$PSD = \frac{N_{RMS}}{\sqrt{BW}}, \quad (3)$$

where $BW$ is the signal bandwidth [56]. The PSD in equation (3) is expressed in units of $\mu g/\sqrt{Hz}$. The calculated PSD values, based on the measured RMS noise in equation (2), were compared to the PSD values in the data sheets of the IMUs. The TreCap accelerometer is operating in low noise mode with an ODR of 250 Hz and a maximum PSD noise value of 260 $\mu g/\sqrt{Hz}$ (Table 1). With a sampling frequency $f_s$ of 200 Hz, an $N_{RMS}$ of 3.7 mg is calculated. The calculated RMS noise does not include the quantisation noise [56].

Several statistics have been extracted from the measured data in rest (noise) and in motion (vibration). For example, peak-to-peak value, absolute peak value, and RMS were calculated in the time domain. The peak-to-peak value indicates the maximum excursion of the vibration wave; it is a useful quantity for measuring the displacement. In analogy, the absolute peak value indicates the maximum level that has occurred. RMS values take the time history of the vibration wave into account and give a value that is directly related to the energy content of the vibration [57]. Estimates of the power spectrum were computed using the squared magnitude of the fast Fourier transform (FFT) algorithm in MATLAB®.

C2. <u>Software validation:</u>
A further aim was to experimentally validate the TreCap software with its functions under real conditions, as shown in Figure 5. The usability of the GUI was evaluated by two independent experts in movement disorders and tested during the different stages of the development process of the device. In our pilot study, movement data was collected from a small cohort of healthy subjects (10 in total, age 60 to 70 years). Physiological hand tremor was measured for all visually guided hand movements to simulate clinical practice. All subjects were seated comfortably in a relaxed environment and each motor task was repeated three times. An evaluation of the software algorithms, for example, cross-validated and compared with more recent studies in the literature, could not be further elaborated due to the novelty of the software. The algorithms were already tested and evaluated with data from our previous clinical studies. In addition, our clinical experience and testing indicate that the research software can be used effectively in a clinical process. To ensure this, we are planning to evaluate the novel software in a clinical study that will be approved by the local ethics committee in accordance with the Declaration of Helsinki.

**D. Software utilisation (example – motor task classifier)**
Machine learning (ML) algorithms were used to distinguish the visually guided hand movements of the healthy subjects as depicted in Figure 5. The proposed approach for motor task classification contains quantification and data labelling (segmentation), filtering and normalisation, and feature extraction to train supervised learning algorithms. Sensor data of the 9-axis IMU were captured with a sampling rate of 200 Hz. Data processing and analysis were conducted on all the 9 sensor axes of the IMU individually. For each motor task, signals with a duration of 5 s were segmented from the mid-position to remove any unstable parts of the signal. The time series data were band-pass filtered (BPF) between 0.25 Hz and 20 Hz to remove artefacts such as drift and noise by using a third-order Butterworth filter. Subsequently, the power spectrum was calculated based on the squared magnitude of the FFT and directly used as a feature for the classification. Each frequency bin of the power spectrum of each segment was divided by the total sum of the frequency bins in the power spectrum of each segment. The normalised power spectrum segments of each axis of the 9-axis IMU signal were concatenated and treated as a continuous spectrum. To capture different speeds of motion and hesitations within the motor tasks more accurately, the short-time power spectrum was calculated based on the squared magnitude of the short-time Fourier transform (STFT). Figure 6 depicts a detailed illustration of the data analysis procedure.

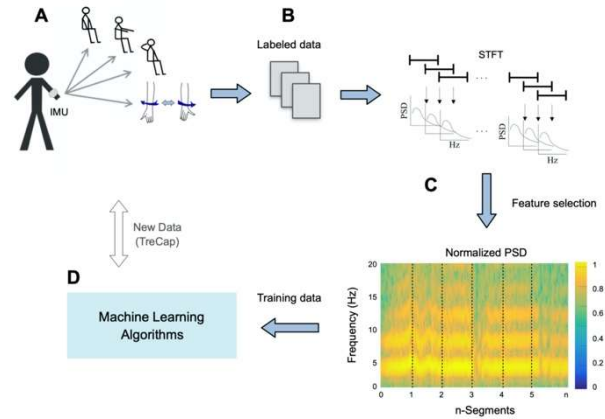

**Figure 6:** Data analysis procedure to extract the features for the motor task classification. The procedure consists of (**A**) data quantification and collection, (**B**) data pre-processing and segmentation (visual labelling) (**C**) feature extraction (selection and tuning), and (**D**) ML classification steps. The normalised power spectrum (range [0,1]) for each short time window was calculated and concatenated to a continuous spectrum. This was applied to each of the 9-axis IMU signals.

In this pilot study, decision tree (DT), discriminant analysis (DA), support vector machine (SVM), and k-nearest-neighbour (kNN) algorithms were explored as base learners [58] for motor task discrimination. Classifier training was performed by using the 5-fold cross-validation procedure to avoid overfitting and to evaluate the classification performance [58].

# Results

Based on our clinical experience, tremor examinations last for one to two hours on average for patients with medication therapy (no DBS), patients with DBS therapy, and patients for DBS surgery screening. Data logging of the TreCap sensor has an autonomy of about 5 hours, including data storage on the microSD card and streaming of kinematic data (with ADC activated) in real-time to the TreCap software without battery recharging. For this duration, the average current consumption of the device was measured with 168 mA, reflecting very small heat power. Three lithium-polymer batteries are available, which can be easily changed. The maximum data transfer rate of the IMU is 144 kpbs. This results in a storage of 0.389 GB on the microSD card for an operating time of 5 hours. Markers from the metadata such as rating scores require less storage space. Based on 100 measurements, the total execution time to process, analyse, and visualise data streams in the MATLAB®-based GUI is 0.467 ms per sample. The transmission speed of the Bluetooth® radio device was tested to be 560 kbps and latency is estimated to be less than 60 ms. TreCap's total latency of execution time, data streaming, converting, storing, and visualisation is estimated to be less than 160 ms. The results of the experimental validation and pilot study are summarised here.

### A. Sensor validation

Sensor noise was measured with the VIW. According to the computed autocorrelation (not illustrated here), TreCap sensor noise can be regarded as an impulse function with high fidelity. This indicates that the acceleration noise can be treated as white noise. Figure 7 shows the noise acceleration with the PSD's. The PSDs are different and depend on the ASIC and MEMS characteristics of the sensors [52]. The PSD of the CS is more uniformly distributed with predominantly higher amplitudes. Figure 8 shows the corresponding box plot and histogram with a normal distribution curve fit, i.e. sensor noise is of the Gaussian type. The results in Table 2 depict that the TreCap sensor has lower noise level compared to the CS for each axis alignment of the g-cell to the Earth's gravitational field component (see legend in Table 2).

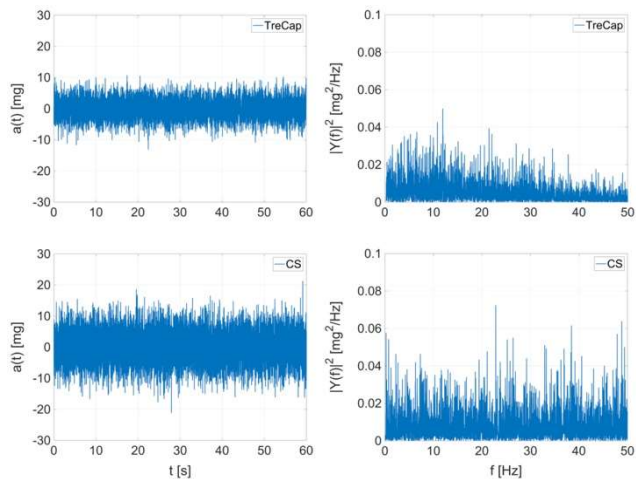

**Figure 7:** Results of measured sensor noise on the VIW (Part I). Sensor alignment of the g-cell in Z-axis. Noise acceleration (left column) with PSD (right column) for TreCap and CS. The PSD figures were computed by using the FFT algorithm in MATLAB®.

For calibration and motion analysis, the vibration exciter Type 4809 was used to ensure that the inertial sensors of TreCap operated according to the specification. In general, IMU sensors provide measurements affected by offsets and drifts. The data sheets of the manufacturers describe the characteristics of these changes. Here, the internal calibration of the IMU was used to perform the experimental validation [46].

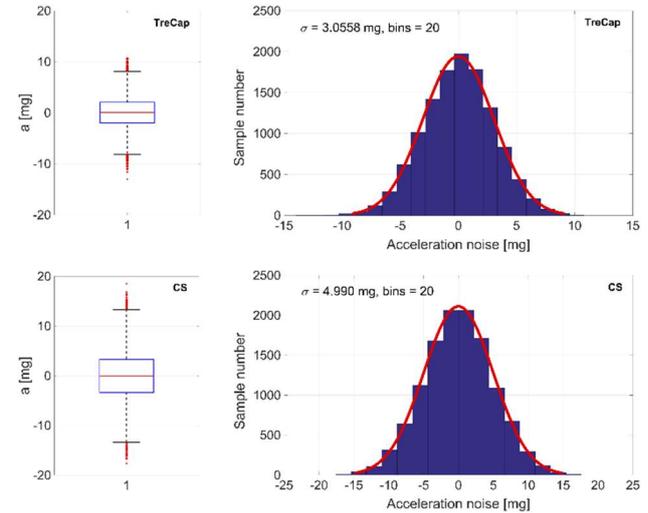

**Figure 8:** Results of measured sensor noise on the VIW (Part II). Sensor alignment with the g-cell in Z-axis. Box plots (left column) and histogram (right column) with normal distribution curve fit (red lines) refer to signals in Figure 7. The spread of the X-axis in the histograms!

**Table 2:** Results of measured sensor noise on the VIW. RMS noise is divided by signal bandwidth in √Hz to determine the PSD noise. The results correspond to the information in Table 1. The IMU of the CS has a PSD noise of 400 µg/√Hz with a measurement range of ±2 g. The ODR is 1 kHz and not adjustable [54]. This results in high values of the g-cell in Z-axis. TreCap was set to an ODR of 250 Hz with a range of ± 2g.

| Accelerometer data | | | | | | |
|---|---|---|---|---|---|---|
| **Metrics** | **RMS (mg)** | | **Peak-to-Peak magnitude (mg)** | | **PSD (µg/√Hz)** | |
| **Sensor** | TC | CS | TC | CS | TC | CS |
| **X as Z** | 2.66 | 3.13 | 22.44 | 25.25 | 188 | 221 |
| **Y as Z** | 2.75 | 3.03 | 22.44 | 23.68 | 195 | 215 |
| **Z as Z** | 3.06 | 4.99 | 23.66 | 42.29 | 216 | 353 |

For motion analysis, the excitation generator was driven to obtain an output displacement with TreCap of about 10 m/s$^2$ (peak-to-peak, about 1.02 g). Figure 9 shows the acceleration with PSD. The sensor prototype is able to detect the excitation frequency with their harmonics. The corresponding results in Table 3 depict that our TreCap sensor is competitive with the CS for each axis alignment of the g-cell on the vibration exciter. We also measured slow vibrations with an output acceleration of 5 m/s$^2$ peak to peak. The results, which are similar to fast vibrations, are not listed here.

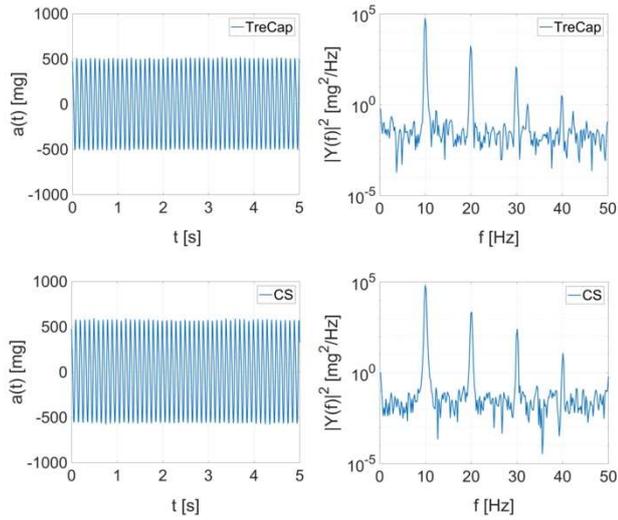

**Figure 9:** Raw sensor data output driven by the vibration exciter. Sensor alignment of the g-cell in Z. Acceleration signal (left column) of TreCap and CS with PSD (right column). The PSD figures were computed using the FFT algorithm in MATLAB®.

**Table 3:** Results of measured sensor motion on the vibration exciter. Sensor axes of the accelerometer were aligned to the Earth's gravitational field component to measure each axis.

| Accelerometer data | | | | | | |
|---|---|---|---|---|---|---|
| Metrics | RMS (mg) | | Peak-to-Peak magnitude (mg) | | PSD (µg/√Hz) | |
| Sensor | TC | CS | TC | CS | TC | CS |
| X as Z | 346 | 392 | 1024 | 1181 | 24.49 | 27.75 |
| Y as Z | 343 | 395 | 1021 | 1184 | 24.26 | 27.93 |
| Z as Z | 347 | 388 | 1031 | 1166 | 24.54 | 27.42 |

**B. Software validation**

In our pilot study, physiological hand tremor was measured for clinically relevant motor tasks and assessed afterwards by a movement disorders expert. In addition, an advanced statistical analysis was performed afterwards. All subjects found it comfortable to wear the TreCap device. The whole test duration for each subject was less than 10 minutes, including set-up, counselling, and rest period, to test each visually guided hand movement at least three times. The user-friendliness of the device was evaluated by two experts in movement disorders based on clinical questions organized in a Likert scale. Physiological hand tremor in posture position (PP) was evaluated with arms extended in front of the body. Intention tremor was tested with the FN motor tasks. Both tasks were evaluated according to the UPDRS motor examination part III. Figure 10 shows an example of the captured data of a healthy subject, as presented in Figure 5, from the pilot study with the corresponding results in Table 4. For the purpose of illustration, the raw data of TreCap in Y-axis for the task PP is illustrated (Figure10, top left), which shows the tremor-dominant sensor axis on the right arm.

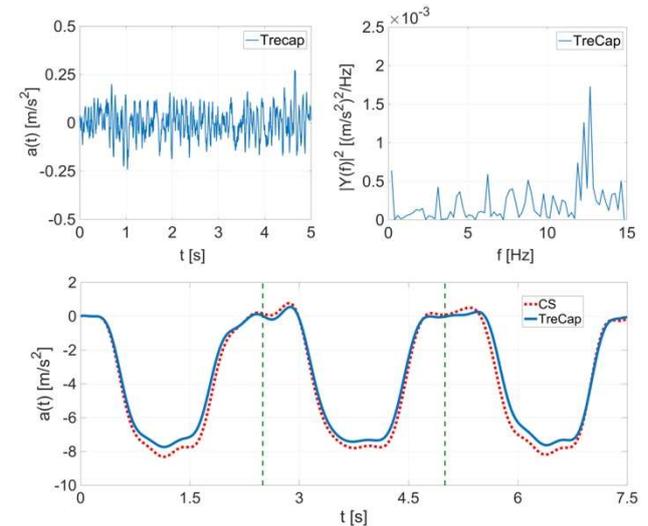

**Figure 10:** Raw Results from the pilot study (excerpt of Subject 2). Raw TreCap data of Y-axis with PSD (peak at 12.7 Hz for PP, top). Physiological hand tremor was effectively quantified by acceleration amplitude (top left) and frequency (top right). FN was performed three times (dashed green lines, bottom). To compare the trajectories of the FN tasks in Y-axis, signals were filtered and aligned to vertical zero-axis.

The results confirm that physiological hand tremor is characterised by low amplitude and high frequency components (Figure 10, top right) as described in the introduction [59], [60]. FN was performed three times (Figure 10, bottom) and the Y-axis of the sensor (i.e. tremor-dominant axis) depicts the distal trajectory on the right arm. To compare FN trajectories between TreCap and the CS, tremor signals were band-pass filtered between 0.25 Hz and 2.25 Hz. Similar results for PP and FN were observed for the X-axis and Z-axis as visualised in Figure 5. This example confirms that the two sensors follow nearly the same trajectory with comparable statistical metrics as in Table 4.

**Table 4:** Measurement results from the pilot study (excerpt). RMS values are equal to the standard deviation. Values for PP and FN refer to the Y-axis (see Figure 10).

| Accelerometer data | | | | | |
|---|---|---|---|---|---|
| Metrics | RMS (mg) | | Peak-to-Peak magnitude (mg) | | PSD (µg/√Hz) |
| Sensor | TC | CS | TC | CS | TC | CS |
| PP | 4.54 | 5.52 | 34.11 | 30.23 | 321 | 391 |
| FN | 325 | 351 | 929 | 988 | 22982 | 24826 |

Moreover, the filtering of physiological tremor is more challenging with the presence of drift, noise, and gravity in acceleration measurements [61], [62]. TreCap's software compensates acceleration noise and gravity, and the motion merging engine of the IMU [46] compensates the acceleration drift. Based on this, the results of repeated measurements of physiological hand tremor show a very high degree of reproducibility in terms of power distribution and frequency, as described in [16], [59].

### C. Classification of motor tasks

For each motor task, 20 data sequences (both hands of each healthy subject) are used for classifier training. The physiological variability in performing the hand movements and differences in performances between the left and right hands were not evaluated and require future investigations. Based on the visual inspection of the hand movement patterns, STFT squared was used with a 3–s window overlapped with a 1–s window as initial estimate, rather than using FFT squared within the entire 5–s signals. This initial estimate is based on prior experience and visual observation of the motor tasks [50]. Table 5 lists the classification performance for both features (FFT squared and STFT squared). The classification results show the overall accuracy of the four motor tasks as depicted in Figure 5.

**Table 5:** Overall accuracies from two feature extraction approaches. Predictors show the length of the input variables (concatenated power spectrum). Model type selection criteria: SVM relies on the one-versus-one method to tackle the multiclass problem with box constraint level of 1; DT with Gini's diversity index and 100 splits (fine); DA with full covariance structure; and kNN with equal distance weight and number of neighbours: 1 (fine) or 10 (cosine), both with Euclidean distance metric. The kernels were selected by an automatic scale.

| Feature selection | Classifiers with accuracy (selected kernel) | | | | Predictors |
|---|---|---|---|---|---|
| | DT | DA | SVM | kNN | |
| FFT squared | 78.8% fine | 91.2 % linear | **91.2 % linear** | 87.5 % cosine | 1828 |
| STFT squared | 73.8 % fine | 93.8 % linear | **95.0 % linear** | 85.0 % fine | 7308 |

The selection of the optimal classifier is a trade-off between feature, algorithm with kernel, and training time, which is mainly dependent on the number of predictors and classes. Table 6 shows the confusion matrix of the classification results for STFT squared with a short time window of 3 s and an overlap of 1 s. The sensitivity depends on true positive (TP) values and false negative (FNE) values (each row) and specificity on true negative (TN) values and false positive (FP) values (each column).

**Table 6:** Classification performance in terms of accuracy, recall and precision. Each row represents the instances in a predicted class while each column represents the instances in a true (actual) class. Kappa was calculated based on [63]. Abbreviations: ROC, Receiver Operating Characteristic; AUC, Area Under Curve; TPR, True Positive Rate; TNR, True Negative Rate; False Negative Rate, FNR (1-TPR); False Positive Rate (1-TNR); PPV, Positive Predictive Value; NPV, Negative Predictive Value, False Discovery Rate (1-PPV); False Omission Rate (1-NPV).

| SVM classifier Healthy subjects' Accuracy 95.0 % | | Predicted class | | | | TPR & TNR | Total sites | Incorrectly classified sites (FNE) | FNR & FPR |
|---|---|---|---|---|---|---|---|---|---|
| | | RP | PP | FN | HM | | | | |
| True class | RP | 17 | 3 | 0 | 0 | 0.85 | 20 | 3 | 0.15 |
| | PP | 1 | 19 | 0 | 0 | 0.95 | 20 | 1 | 0.05 |
| | FN | 0 | 0 | 20 | 0 | 1 | 20 | 0 | 0 |
| | HM | 0 | 0 | 0 | 20 | 1 | 20 | 0 | 0 |
| PPV & NPV | | 0.94 | 0.86 | 1 | 1 | | | | |
| Total classified sites | | 18 | 22 | 20 | 20 | ROC curve parameters: Class 1: AUC = 0.97, FPR = 0.02; Class 2: AUC = 0.97, FPR = 0.05; Class 3: AUC = 1.00, FPR = 0; Class 4: AUC = 1.00, FPR = 0; Kappa = 0.9333 | | | |
| Incorrectly classified sites (FP) | | 1 | 3 | 0 | 0 | | | | |
| (FDR & FOR) | | 0.056 | 0.14 | 0 | 0 | | | | |

The classification performance of each motor task is shown in Table 7. Based on the multiclass classification problem, some of the classification performance parameters are equal, since TP, TN, FP and FNE are normally encountered in binary classifiers (yes/no decisions).

**Table 7:** Classification performance metrics of each motor task. The average of the accuracies of all the classes is 97.5 % and should not be confused with the overall accuracy in Table 6 (95 %). TPR (Sensitivity) is also a parameter of the ROC curve and depicted separately. Additionally, FPR may differ from the equation (1 – Specificity), since this parameter was extracted from the ROC curve which shows the optimal classifier performance in each class. Abbreviations: Matthews Correlation Coefficient, MCC.

| Metrics | RP | PP | FN | HM |
|---|---|---|---|---|
| TP | 17 | 19 | 20 | 20 |
| TN | 59 | 57 | 56 | 56 |
| FP | 1 | 3 | 0 | 0 |
| FNE | 3 | 0 | 0 | 0 |
| Accuracy | 0.950 | 0.950 | 1 | 1 |
| Precision (PPV) | 0.944 | 0.863 | 1 | 1 |
| Sensitivity (TPR) | 0.850 | 0.950 | 1 | 1 |
| Specificity (TNR) | 0.983 | 0.950 | 1 | 1 |
| NPV | 0.951 | 1 | 1 | 1 |
| F-score | 0.894 | 0.904 | 1 | 1 |
| MCC | 0.864 | 0.906 | 1 | 1 |

# Discussion

Several commercial products of wearable devices are used for medical research, clinical rehabilitation, ergonomics and sports science (e.g. Biometrics[TM] Ltd., UK; APDM[TM] Inc., USA; and Shimmer[TM] Ltd., Ireland). The software of these devices, however, does not address specific requirements in a clinical setting. A counter example of mobile health technology is the Parkinson's KinetiGraph[TM] System (PKG™, Global Kinetics Corp., Australia), which is a wrist-worn movement recording device for patients with PD. This commercial device shares similar aims with TreCap. However, the PKG™ system is mainly designed to track daily motor fluctuations in PD outside the clinic with denied access to raw data and pre-processed algorithms for data visualisation. In contrast, TreCap is designed to assist clinical researchers who frequently encounter movement disorder patients. TreCap, with sensor performances comparable to the CS reference, provides novel software features such as metadata logging (side effects, DBS parameters, and more), one-click scoring, and further customisations.

All sensors of the TreCap device were technically validated. The functionality of the software was validated by measuring the physiological hand tremor of an elderly population. Two experts for movement disorders evaluated the GUI. So far, the validation of our software with other systems specified in the literature is still open, since most of the functions are new. This work, however, does not cover how far the use of TreCap can ease the clinical interpretation of the severity of motor symptoms and support the clinical decision-making process. Hence, we are planning to evaluate TreCap in a clinical study. The effectiveness of the system in supporting clinical diagnosis and treatment is a long-term goal of our research and essentially depends on the evaluation and analysis of the patients' pathological movement patterns [64], [65].

The TreCap software suite offers the platform to implement classifiers for motor task detection, an objective assessment of motor symptoms and to collect new data for classifier training. Based on the pilot study in this work, the next step is to apply the trained classification model in real time to automatically identify, assess and process the motor tasks of patients. This requires an optimisation of the classifier (embedding patient data, updated feature settings and algorithms) to identify whether a segment of ordinary movements shows early signs of tremor. We demonstrate the classification of relevant hand motor tasks within one examination by using a single IMU. This approach has not yet been employed.

Finally, TreCap could provide a platform suitable for advanced clinical trials in movement disorders. Future examples of applications may include classifiers to diagnose various types of motor symptoms and to predict their severity [64], [65], [66]. This is particularly relevant to predict the transition from physiological to pathological tremor (early warning), as already observed in individuals at high risk for familial ET [16], [67]. Hence, large clinical studies are warranted to determine the diagnostic accuracy for each motor symptom [42], [67].

# Conclusion

This work presents TreCap, a wearable device with a novel software to evaluate and manage tremor data of visually guided hand movements in real time. The system is developed for clinical researchers who frequently encounter movement disorder patients. Inertial sensors on the PCB measure limb kinematics and a bioelectrical unit is integrated to measure neuromuscular activity. Sensor data is transmitted via a Bluetooth® radio device to a custom-written GUI in MATLAB®. The microcontroller on the PCB is capable of executing algorithms under several conditions and sensor data are optionally stored on a microSD card. TreCap's software contains new functions to label and process motor task data within a short rework time.

Our focus will be on the further development and clinical validation of the system. TreCap offers clinical researchers in movement disorders the opportunity to work in a clinical setting according to a more objective and standardised procedure. Hardware, firmware and software of the wearable device have been verified on the basis of our experimental validation and pilot study. Finally, TreCap has the potential to serve as a medical assistance system for diagnostic workups and therapeutic decisions.

# Authors statement

This work was supported by the Luxembourg National Research Fund (FNR, AFR-PhD Grant 10086156, Rene Peter Bremm). The authors have no conflict of interest to declare. Informed consent has been obtained from all healthy subjects in the pilot study (no personal data, sensor data anonymised) has received a favorable opinion from the National Research Ethics Committee (CNER) Luxembourg (N.201703/01, OptiStimDBS). The work on TreCap is embedded in the collaboration between the Systems Control Group and Interventional Neuroscience Group (LCSB, Luxembourg Centre for Systems Biomedicine, University of Luxembourg, Esch-sur-Alzette, Luxembourg), National Department of Neurosurgery (CHL, Centre Hospitalier de Luxembourg, Luxembourg) and Department of Electrical Engineering (TUAS, Trier University of Applied Sciences, Trier, Germany). The authors would like to thank for the support of Michael Hoffmann, M. Sc. (TUAS, Department of Mechanical Engineering), who aided in designing and manufacturing the sensor housing via 3D rapid prototyping technology. Furthermore, the authors would like to thank Prof. Dr. Rejko Krüger (LCSB, Translational Neuroscience Group), and Dr. Christophe Berthold (CHL, National Department of Neurosurgery) for their clinical expertise.

# References

[1] L. Lonini et al., "Wearable sensors for Parkinson's disease: which data are worth collecting for training symptom detection models," *npj Digit. Med.*, vol. 1, no. 64, pp. 8 pages, Nov. 2018.

[2] A. D. Trister et al., "Smartphones as new tools in the management and understanding of Parkinson's disease," *npj Park. Dis.*, vol. 3, no. 2, 2 pages, Mar. 2016.

[3] D. J. Wile et al., "Smart watch accelerometry for analysis and diagnosis of tremor," *J. Neurosci. Methods*, vol. 230, 4 pages, June 2014.

[4] A. T. Tzallas et al., "Perform: A system for monitoring, Assessment and management of patients with Parkinson's disease," *Sensors (Switzerland)*, vol. 14, no. 11, pp. 21329-21357, Nov. 2014.

[5] E. R. Dorsey et al., "Projected number of people with Parkinson disease in the most populous nations, 2005 through 2030," *Neurology,* vol. 69, no. 2, pp. 223-224, July 2007.

[6] P. G. Bain, "The management of tremor," *J. Neurol. Neurosurg. Psychiatry*, vol. 1, no. 1, pp. 13-19, Apr. 2002.

[7] P. Odin et al., "Viewpoint and practical recommendations from a movement disorder specialist panel on objective measurement in the clinical management of Parkinson's disease," *npj Park. Dis.*, vol. 4, no. 14, 7 pages, May 2018.

[8] D. Charvin et al., "Therapeutic strategies for Parkinson disease: Beyond dopaminergic drugs," *Nature Rev. Drug Discovery*, vol. 17, no. 11, 19 pages, Nov. 2018.

[9] K. Ashkan et al., "Insights into the mechanisms of deep brain stimulation," *Nature Rev. Neurol.* vol. 13, no. 9, pp. 548-554 Sep. 2017.

[10] W. M. M. Schuepbach et al., "Neurostimulation for Parkinson's Disease with Early Motor Complications," *N. Engl. J. Med.*, vol. 368, no. 7, pp. 610–622, Feb. 2013.

[11] C. G. Goetz et al., "Movement Disorder Society-sponsored revision of the Unified Parkinson's Disease Rating Scale (MDS-UPDRS): Process, format, and clinimetric testing plan," *Mov. Disord.*, vol. 22, no. 1, pp. 41–47, Jan. 2007.

[12] R. Elble et al., "The essential tremor rating assessment scale (TETRAS)," *J Neurol Neuromed.*, vol. 1, no. 4, pp. 34–38, 2016.

[13] C. Ramaker et al., "Systematic evaluation of rating scales for impairment and disability in Parkinson's disease," *Mov. Disord.*, vol. 17, no. 5, pp. 867-876, Sep. 2002.

[14] J. H. McAuley, "Physiological and pathological tremors and rhythmic central motor control," *Brain*, vol. 123, no. 8, pp. 1545-1567, Aug. 2000.

[15] J. Timmer et al., "Quantitative analysis of tremor time series," *Electroencephalogr. & Clin. Neurophysiol.*, vol. 101, no. 5, pp. 461–468, Mar. 1996.

[16] R. J. Elble, "Characteristics of physiologic tremor in young and elderly adults," *Clin. Neurophysiol.*, vol. 114, no. 4, pp. 624-635, Dec. 2003.

[17] J. Raethjen et al., "Tremor analysis in two normal cohorts," *Clin. Neurophysiol.*, vol. 115, no. 9, pp. 2151-2156, Sep. 2004.

[18] S. H. Alusi, "A study of tremor in multiple sclerosis," *Brain*, vol. 124, no. 4, pp. 720-730, April 2001.

[19] J. Jankovic, "Essential tremor: clinical characteristics," *Neurology.*, vol. 54, no. 11, Suppl. 4, pp. 21-25, Feb. 2000.

[20] E. Dietrichs, "Clinical manifestation of focal cerebellar disease as related to the organization of neural pathways," *Acta Neurologica Scand.*, vol. 117, no. s188, pp. 6-11, Apr. 2008.

[21] K. Bötzel, et al., "The Differential Diagnosis and Treatment of Tremor," *Dtsch. Aerzteblatt Int.*, vol. 111, no. 13, pp. 225-236, Mar. 2014.

[22] T. R. Bennett et al., "Inertial measurement unit-based wearable computers for assisted living applications: A signal processing perspective," *IEEE Signal Process. Mag.*, vol. 33, no. 2, pp. 28–35, March 2016.

[23] W. Maetzler et al., "A clinical view on the development of technology-based tools in managing Parkinson's disease," *Mov. Disord.*, vol. 31, no. 9, pp. 1263-1271, Sep. 2016.

[24] D. G. M. Zwartjes et al., "Ambulatory monitoring of activities and motor symptoms in Parkinson's disease," *IEEE Trans. Biomed. Eng.*, vol. 57, no. 11, pp. 2778-2786, Nor. 2010.

[25] L. Allet et al., "Wearable systems for monitoring mobility-related activities in chronic disease: A systematic review," *Sensors (Switzerland)*, vol. 10, no. 10, pp. 9026-9052, Oct. 2010.

[26] J. F. Daneault et al., "Exploring the use of wearable sensors to monitor drug response of patients with Parkinson's disease in the home setting," in J. *Neurol.*, vol. 88, Suppl. 16, April 2017.

[27] D. Johansson *et al.*, "Wearable sensors for clinical applications in epilepsy, Parkinson's disease, and stroke: a mixed-methods systematic review," *J. Neurol.*, vol. 265, no. 8, pp. 1740-1752, Aug. 2018.

[28] C. L. Pulliam *et al.*, "Continuous assessment of levodopa response in Parkinson's disease using wearable motion sensors," *IEEE Trans. Biomed. Eng.*, vol. 65, no. 1, pp. 159-164, Jan. 2018.

[29] Á. Sánchez-Ferro *et al.*, "New methods for the assessment of Parkinson's disease (2005 to 2015): A systematic review," *Mov. Disord.*, vol. 31, no. 9, pp. 1283-1292, July 2016.

[30] J. Klucken *et al.*, "Unbiased and Mobile Gait Analysis Detects Motor Impairment in Parkinson's Disease," *PLoS One*, vol. 8, no. 2, 9 pages, Feb. 2013.

[31] N. Mahadevan *et al.*, "Development of digital biomarkers for resting tremor and bradykinesia using a wrist-worn wearable device," *NPJ Digital Med..*, vol. 3, no. 5, pp. 1–12, Jan. 2020.

[32] E. Rovini, *et al.*, "How wearable sensors can support Parkinson's disease diagnosis and treatment: A systematic review," *Front. Neurosci.*, vol. 11, Article 555, 41 pages, Oct. 2017.

[33] S. Shahtalebi *et al.*, "PHTNet: Characterization and Deep Mining of Involuntary Pathological Hand Tremor using Recurrent Neural Network Models," *Science Reports*, vol. 10, Article 2915, 7 pages, Feb. 2020.

[34] A. Maldonado-Naranjo *et al.*, "Kinematic Metrics from a Wireless Stylus Quantify Tremor and Bradykinesia in Parkinson's disease," *Parkinsons Dis.*, vol. 2019, no. 1, 9 pages, April 2019.

[35] G. Grimaldi and M. Manto, "Neurological tremor: Sensors, signal processing and emerging applications," *Sensors (Switzerland)*, vol. 10, no. 2, pp. 1399–1422, Feb. 2010.

[36] D. Lukšys *et al.*, "Quantitative Analysis of Parkinsonian Tremor in a Clinical Setting Using Inertial Measurement Units," *Parkinsons. Dis.*, vol. 2018, Article ID 1683831, 7 pages, June 2018.

[37] G. J. Welk, "Principles of design and analyses for the calibration of accelerometry-based activity monitors," in *Med. and Science in Sports and Exercise*, vol. 37, no s11, pp. 501-511, Nov. 2005.

[38] A. M. Sabatini, "Estimating three-dimensional orientation of human body parts by inertial/magnetic sensing," *Sensors (Switzerland)*, vol. 11, no. 2, pp. 1489-1525, Jan. 2011.

[39] V. Ruonala *et al.*, "EMG signal morphology and kinematic parameters in essential tremor and Parkinson's disease patients," *J. Electromyogr. Kinesiol.*, vol. 24, no. 2, pp. 300–306, Apr. 2014.

[40] Zhang *et al.*, "Differential diagnosis of Parkinson's disease, essential tremor, and enhanced physiological tremor with the tremor analysis of EMG," *Parkinsons. Dis.*, vol. 2017, Article ID 1597907, 4 pages, Aug. 2017.

[41] P. Farzanehfar *et al.*, "Objective measurement in routine care of people with Parkinson's disease improves outcomes," *npj Park. Dis.*, vol. 4, 7 pages, March 2018.

[42] H. C. Powell *et al.*, "A wearable inertial sensing technology for clinical assessment of tremor," in *Conference Proceedings - IEEE Biomedical Circuits and Systems, BiOCAS2007*, pp. 9-12, 2007.

[43] S. Spasojević *et al.*, "Quantitative assessment of the arm/hand movements in Parkinson's disease using a wireless armband device," *Front. Neurol.*, vol. 8, 15 pages, Aug. 2017.

[44] R. P. Bremm, J. Gonçalves, K. P. Koch, and F. Hertel, "TreCap – Quantify and assess tremor in real-time with a new wearable device," *Mov. Disord.*, vol. 33, Suppl. 2, 1 page (conference abstract), Oct. 2018.

[45] B. P. Bezruchko and D. A. Smirnov, "Brain – Limb Couplings in Parkinsonian Resting Tremor," Chapter 12.3, Page 327, in *Extracting Knowledge from Time Series: An Introduction to Nonlinear Empirical Modeling*, ed. 2010, Springer (New York), 410 pages, 2010.

[46] Maxim Integrated Corp., MAX21100 (data sheet), Note 19-6741, Rev 1, 10-2014, Accessed: 22 October 2020, https://www.maximintegrated.com/en/products/interface/sensor-interface/MAX21100.html

[47] STMicroelectronics Corp., LIS3MDL (data sheet), Note 024204, Rev 6, 05-2017, Accessed: 22 October 2020, https://www.st.com/en/mems-and-sensors/lis3mdl.html

[48] A. Amon and F. Alesch, "Systems for deep brain stimulation: review of technical features," *J. Neural Transm.*, vol. 124, no. 9, pp. 1083–1091, Sep. 2017.

[49] A. Bruns, "Fourier-, Hilbert- and wavelet-based signal analysis: Are they really different approaches?," *J. Neurosci. Methods*, vol. 137, no. 2, pp. 321-332, Aug. 2004.

[50] R. P. Bremm, K. P. Koch, R. Krüger, J. Gonçalves, and F. Hertel, "Analysis and visualisation of tremor dynamics in deep brain stimulation patients," in *Biomed Eng. (Biomed Tech.): 54th Annual Conference of the German Society for Biomedical Engineering*, Leipzig, Germany, 29 Sep. – 1 Oct. 2020, 4 pages.

[51] R. P. Bremm, K. P. Koch, R. Krüger, F. Hertel, and J. Gonçalves, "A rule-based expert system for real-time feedback-control in deep brain stimulation," in *Biomed Eng. (Biomed Tech.): 54th Annual Conference of the German Society for Biomedical Engineering*, Leipzig, Germany, 29 Sep. – 1 Oct. 2020, 4 pages.

[52] E. Rovini, C. Maremmani, and F. Cavallo, "How wearable sensors can support parkinson's disease diagnosis and treatment: A systematic review," *Front. Neurosci.*, vol. 11, no. 555, pp. 1–41, Oct. 2017.

[53] P. Blank *et al.*, "miPod 2: A new Hardware Platform for Embedded Real-Time Processing in Sports and Fitness Applications," in *Conference Proceedings - Ubiquitous Computing Adjunct 2016*.

[54] InvenSense Corp., MPU-9150 (data sheet), Note PS-MPU-9150A-00, Rev 4.3, 09-2013, https://www.invensense.com, Accessed: 1 Aug. 2019.

[55] B. Reeder *et al.*, "Health at hand: A systematic review of smart watch users for health and wellness," *J Biomed Inform.*, vol. 63, no. 1, pp. 269-276, Oct. 2016.

[56] Freescale Semiconductor Inc., Application Note AN4075, Rev 1, 09-2010, Accessed: 22 October 2020, https://www.nxp.com/docs/en/application-note/AN4075.pdf

[57] A. Brandt, *Noise and Vibration Analysis: Signal Analysis and Experimental* Procedures, ed. 2011, Wiley (UK), 464 pages, 2011.

[58] J. D. Kelleher, B. Mac Namee, and A. D'Arcy, *Fundamentals of Machine Learning for Predictive Data Analysis.*, 1st edition, MIT Press, 624 pages, 2015.

[59] G. Grimaldi and M. Manto, *Mechanisms and emerging therapies in tremor disorders*, 1st edition, Springer (New York), 492 pages, 2013.

[60] M. S. Titcombe, R. Edwards, and A. Beuter, "Mathematical modeling of Parkinson's Tremor," *Nonlinear Studies*, vol. 11, no. 3, pp. 363–384, 2004.

[61] K. C. Veluvolu and W. T. Ang, "Estimation of physiological tremor from accelerometers for real-time applications," *Sensors*, vol. 11, no. 3, pp. 3020–3036, March 2011.

[62] C. N. Riviere *et al.*, "Adaptive canceling of physiological tremor for improved precision in microsurgery," *IEEE Trans. Biomed. Eng.*, vol. 45, no. 7, pp. 839-846, July 1998.

[63] J. R. Landis and G. G. Koch, "The Measurement of Observer Agreement for Categorical Data," *Biometrics*, vol. 33, no. 1, pp. 159–174, March 1977.

[64] B. T. Cole *et al.*, "Dynamical learning and tracking of tremor and dyskinesia from wearable sensors," *IEEE Trans. Neural Syst. Rehabil. Eng.*, vol. 22, no. 5, pp. 982-991, Sep. 2014.

[65] C. Gao *et al.*, "Model-based and model-free machine learning techniques for diagnostic prediction and classification of clinical outcomes in Parkinson's disease," *Sci. Rep.*, vol. 8, no. 7129, 21 pages, May 2018.

[66] H. Jeon, W. Lee, H. Park, H. J. Lee, S. K. Kim, H. B. Kim, B. Jeon, K. S. Park, "Automatic classification of tremor severity in Parkinson's disease using a wearable device," *Sensors (Switzerland)*, vol. 17, no. 9, pp. 1–14, Sep. 2017.

[67] R. J. Elble, C. Higgins, and S. Elble, "Electrophysiologic transition from physiologic tremor to essential tremor," *Mov. Disord.*, vol. 20, no. 8, pp. 1038–42, Aug. 2005.